# A High Availability Clusters Model Combined with Load Balancing and Shared Storage Technologies for Web Servers

A B M Moniruzzaman, Md. Waliullah, Md. Sadekur Rahman

**Abstract**— This paper designs and implements a high availability clusters and incorporated with load balance infrastructure of web servers. The paper described system can provide full facilities to the website hosting provider and large business organizations. This system can provide continuous service though any system components fail uncertainly with the help of Linux Virtual Server (LVS) load-balancing cluster technology and combined with virtualization as well as shared storage technology to achieve the three-tier architecture of Web server clusters. This technology not only improves availability, but also affects the security and performance of the application services being requested. Benefits of the system include node failover overcome; network failover overcome; storage limitation overcome and load distribution.

—————————— ◆ ——————————

## 1 INTRODUCTION

THIS is a system which can provide full facilities to the website hosting provider and large business organizations High Availability Cluster with Load Balancing" is such a system, which can provide continuous service though any system components fail uncertainly. Without clustering, a server running a particular application as when that particular server crashes, the application will be unavailable until the crashed server is fixed. This system is very much efficient, because the full system is automated from the very beginning to the end.

High availability is essential for any organizations interested in protecting their business against the risk of a system outage, loss of transactional data, incomplete data, or message processing errors. Servers don't run forever. Hardware components can fail. Software can crash. Systems are shutdown for upgrades and maintenance. Whatever the reason, when a server goes down, the applications and the business processes that depend on those applications stop. For a business interested in being available at all times, HA clustering is a practical solution. High availability clusters allow the application and business process to resume operations quickly despite the failure of a server and ensure business is not interrupted. High availability clusters are simple in principle. Two or more servers are joined or clustered together to back each other up. If the primary server goes down, the clustering system restarts the application on one of the other servers in the cluster, allowing the business to continue operating normally. The servers are connected using a network or serial interface so they can communicate with each other. With this kind of clustering there is no need to modify the application. Companies have looked for ways to deliver HA results, including downloading high availability software solutions, custom development solutions, and even load balancers.

## 2 RELATED TECHNOLOGIES AND ARCHITECTURES

### 2.1 Computer Cluster

A computer cluster consists of a set of connected computers that work together so that in many respects they can be viewed as a single system [1]. The components of a cluster are usually connected to each other through fast local area networks (LAN), with each node (computer used as a server) running its own instance of an operating system. Clusters are usually deployed to improve performance and availability over that of a single computer, while typically being much more cost-effective than single computers of comparable speed or availability [2]. Computer clusters emerged as a result of convergence of a number of computing trends including the availability of low cost microprocessors, high speed networks, and software for high performance distributed computing [1]. Clusters concept is actually virtually management of several highly connected physical machines by combining these standalone hosts into a single entity. Clusters give the power of multiple hosts with the simplicity of managing a single entity with pooled resources and inherently higher availability.

### 2.2 High Availability

High availability (HA) is actually meant by eliminate the scheduled and unscheduled downtime of any computer systems. Typically, scheduled downtime is a result of maintenance that is disruptive to system operation and usually cannot be avoided with a currently installed system design. Scheduled downtime events might include patches to system software that require a reboot or system configuration changes that only take effect upon a reboot.
Unscheduled downtime events typically arise from some physical event, such as a hardware or software failure or environmental anomaly. Examples of unscheduled downtime events include power outages, failed CPU or RAM compo-nents (or possibly other failed hardware components), an over-temperature related shutdown, logically or physically severed network connections, security breaches, or various application, middleware, and operating system failures. To eliminate any single point of failure need carefully implemented specialty designs that allow online hardware, network, operating system, middleware, and application upgrades, patches, and replacements. This paper describes High-availability cluster (HA cluster) implementations which attempt to use redundancy of cluster components to eliminate single points of failure.

### 2.3 High-availability clusters

High-availability clusters also known as HA clusters or failover clusters are groups of computers that support server applications that can be reliably utilized with a minimum of down-time [5]. They operate by harnessing redundant computers in groups or clusters that provide continued service when system components fail. Without clustering, if a server running a particular application crashes, the application will be unavailable until the crashed server is fixed. HA clustering remedies this situation by detecting hardware/software faults, and immediately restarting the application on another system without requiring administrative intervention, a process known as failover. As part of this process, clustering software may configure the node before starting the application on it. For example, appropriate file systems may need to be imported and mounted, network hardware may have to be configured, and some supporting applications may need to be running as well. HA clusters are often used for critical databases, file sharing on a network, business applications, and customer services such as electronic commerce websites. HA cluster implementations attempt to build redundancy into a cluster to eliminate single points of failure, including multiple network connections and data storage which is redundantly connected via storage area networks.

### 2.3 Load Balance

Load balancing is a computer networking method for distributing workloads across multiple computing resources, such as computers, a computer cluster, network links, central processing units or disk drives [8]. Load balancing aims to optimize resource use, maximize throughput, minimize response time, and avoid overload of any one of the resources [9]. Using multiple components with load balancing instead of

- **A B M Moniruzzaman,** Received his B.Sc (Hon's) degree in Computing and Information System (CIS) from London Metropolitan University, London, UK and M.Sc degree in Computer Science and Engineering (CSE) from Daffodil International University, Dhaka, Bangladesh in 2005 and 2013, respectively. Currently he is working as a Lecturer of the department of Computer Science and Engineering ad Daffodil International University. E-mail: monir.cse@daffodilvarsity.edu.bd
- **Md. Waliullah**, Received his B.Sc degree in Computer Science and Engineeringfrom Hajee Mohammad Danesh Science & Technology University and M.Sc. in Network & Computer Systems Security from University of Greenwich, London, UK. Currently he is working as a Lecturer of the department of Computer Science and Engineering ad Daffodil International University. E-mail: waliullah.cse@daffodilvarsity.edu.bd
- **Md. Sadekur Rahman**, Received his B.Sc. & M.Sc. in Applied Mathematics & Informatics from Peoples' Friendship University of Russia. Currently he is working as a Lecturer of the department of Computer Science and Engineering ad Daffodil International University. E-mail: sadekur.cse@daffodilvarsity.edu.bd

a single component may increase reliability through redundancy. Load balancing is usually provided by dedicated software or hardware, such as a multilayer switch or a Domain Name System server process. There are several methods of load balancing like Round-Robin algorithm approach. A Storage Area Network (SAN) is a dedicated network that provides access to consolidated, block level data storage [10].

### 2.4 Failover

In computing, Failover is switching to a redundant or standby computer server, system, hardware component or network upon the failure or abnormal termination of the previously active application, server, system, hardware component, or network [11]. Failover and switchover are essentially the same operation, except that failover is automatic and usually operates without warning, while switchover requires human intervention. Systems designers usually provide failover capability in servers, systems or networks requiring continuous availability. The used term is High Availability- and a high degree of reliability.

### 2.5 Linux Virtual Server (LVS) and virtualization technology

Different cluster usually divided into three basic types [3]: High-Performance Clusters, High-Availability Cluster and Load Balance Cluster. Linux Virtual Server (LVS) [15],[18],[19],[20],[21] in this load balancing cluster technology cluster structure, the server cluster of horizontal expansion. The major work of the LVS project is now to develop advanced IP load balancing software (IPVS), application-level load balancing software (KTCPVS), and cluster management components [16]. IPVS is an advanced IP load balancing software implemented inside the Linux kernel [16]. The IP Virtual Server code was already included into versions 2.4.x and newer of the Linux kernel mainline. KTCPVS implements application-level load balancing inside the Linux kernel, currently under development.

### 2.6 Failover

LVS is mainly deal with four layers of the OSI model information package, according to certain rules to forward the request directly to the back-end services processing nodes. Virtual Server is a server Load Balancer and the logical combination collectively, need only use the service to interact with the Virtual Server can get efficient service. LVS cluster is proportional to the number provided with the server load capacity, the use of IP-based layer to IP layer load balancing of TCP / IP requests to a server in the pool even different servers; When a cluster system failure , the cluster software to respond quickly, the system dynamically allocates tasks to the work of others in the cluster is performed on a system, thus ensuring continuity of services; when the overall work flow beyond the capabilities of each system cluster, it will there are other systems added to the cluster, so that the overall system performance to smoothly expand, and the client are not affected, nor do they need any changes to the external service and efficient [4].

Demand for different web services, LVS scheduler to achieve 10 kinds of load scheduling algorithms, the following four common algorithms: Round Robin, Weighted Round Robin, Least Connections, Weighted Least Connections [17]

## 3 PROPOSED MODEL DESIGN AND IMPLEMENTATION DETAILS

High Availability Cluster with Load Balancing" is such a system, which can provide continuous service though any system components fail uncertainly. Without clustering, a server running a particular application as when that particular server crashes, the application will be unavailable until the crashed server is fixed. This system is very much efficient, because the full system is automated from the very beginning to the end. One of the issues in designing a cluster is how tightly coupled the individual nodes may be. For instance, a single computer job may require frequent communication among nodes: this implies that the cluster shares a dedicated network, is densely located, and probably has homogenous nodes.

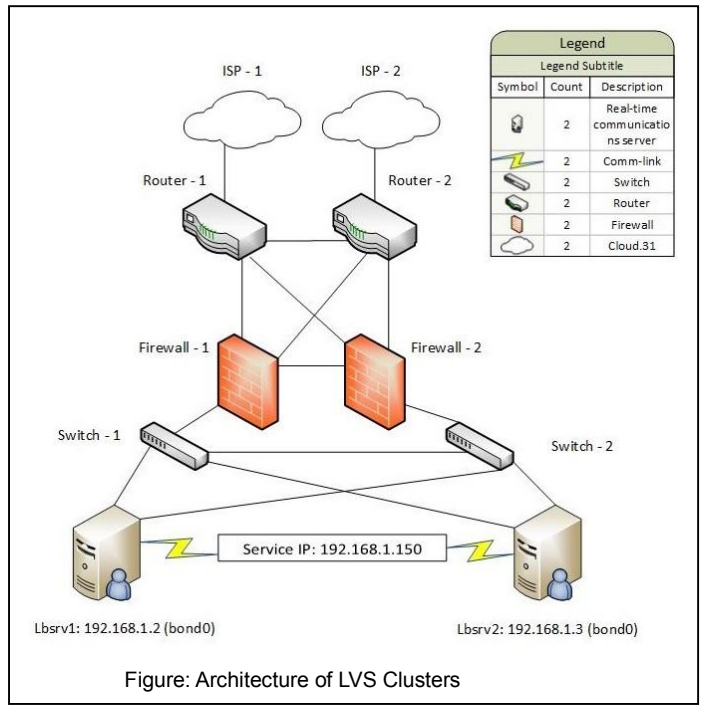

Figure: Architecture of LVS Clusters

In the figure 3.1 the load balancers make parallel services of the cluster to appear as a virtual service on a single IP address (virtual IP), and request dispatching can use IP load balancing technologies

In the figure 3.2 real servers and the load balancers are interconnected with high-speed LAN (Private Network). The load balancers dispatch requests to the different servers and make parallel services of the cluster to appear as a virtual service on a single IP address, and request dispatching can use IP load balancing technologies or application-level load balancing technologies.

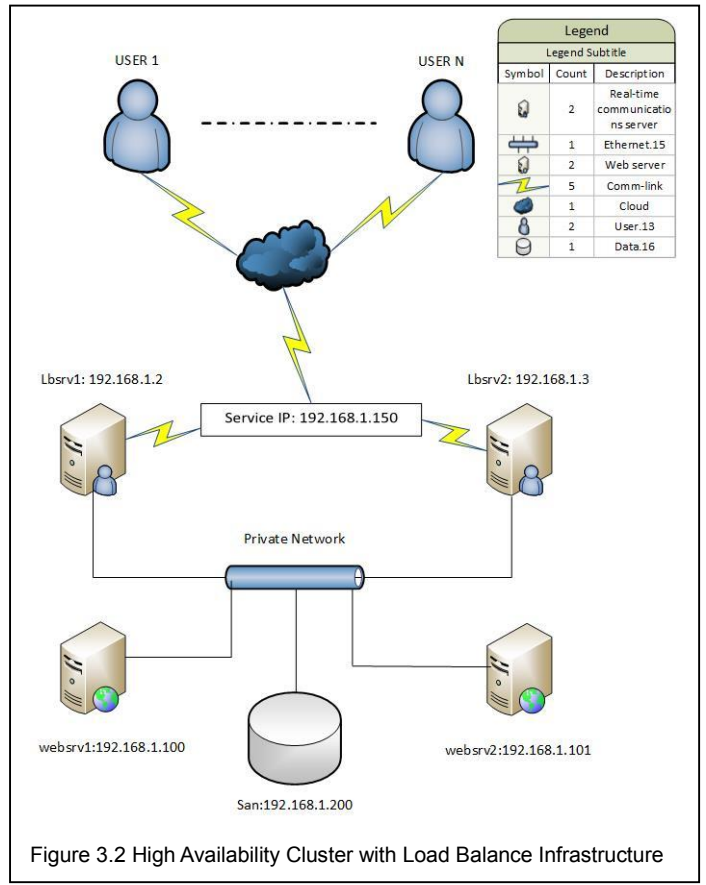

Figure 3.2 High Availability Cluster with Load Balance Infrastructure

Scalability of the system is achieved by transparently adding or removing nodes in the cluster. High availability is provided by detecting node or daemon failures and reconfiguring the system appropriately. In the figure 3.2 all request will be process by Round Robin Algorithm.

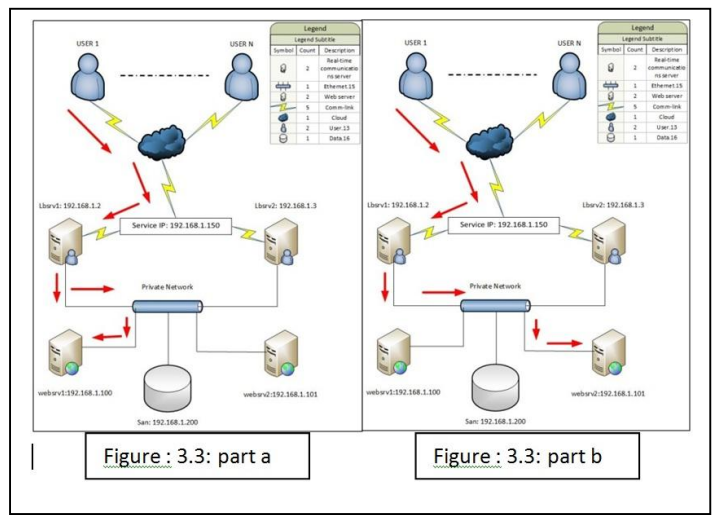

Figure : 3.3: part a    Figure : 3.3: part b

In the example figure 3.3 a,b,c,d parts 1st request will handle websrv1 and 2nd request will be handle websrv2 and again 3rd request will handle websrv1 and process will be go on. Most important thing is all request will accept lbsrv1 until its goes down, when it will goes down then lbsrv2 will take over all the service and it will handle all the requests. Heartbeat network protocol is used this works to monitor.

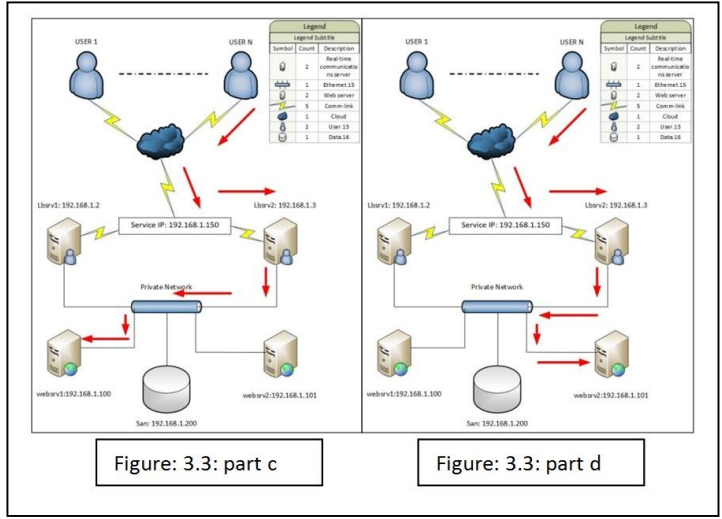

Figure: 3.3: part c    Figure: 3.3: part d

## 4 EXPERIMENTAL SETUP AND TESTS

In this chapter we will discuss about implementation features of this project. The features described with necessary figures. We will configure our computers or nodes as following:

| SL | Host Name | IP | Description |
|----|-----------|-----|-------------|
| 01 | lbnode1.orange.com | 192.168.1.2 | Primary Cluster Server |
| 02 | lbnode2.orange.com | 192.168.1.3 | Secondary Cluster Server |
| 03 | websrv1.orange.com | 192.168.1.100 | Primary Web Server (Active Router) |
| 04 | websrv2.orange.com | 192.168.1.101 | Secondary Web Server (Backup Router) |
| 05 | san.example.com | 192.168.1.200 | Storage Area Network |

Here load balancing will be done using 2 Linux Server. Also have to install two Web servers for surfing web content. There also need SAN to avoid storage problem. Whole system will run under VMware Workstation 8.0 environment.

### 4.2 Custom Red Hat Enterprise Linux" Installation

First of all stop all the services that don't need to run on those nodes. Then, we will modify our host's configuration file at /etc/hosts on each of the nodes in our setup; then we copy the /etc/hosts file to all the servers (This step is not required if you have DNS). After copying to host file to all the nodes, we need to generate SSH keys. Now copy ssh keys to all other nodes for password less entry which is required by pulse daemon. We can build up a global finger print list. Now we will configure NTP service on all the nodes. We will make the LBNODE1 as our NTP Server. Now we will configure client side configuration in WEBSRV1. Copy the same configuration or the file /etc/ntp.conf to other 2 nodes websrv2, lbnode2. After copying restart the ntp service on these nodes. Now we will update the time on all the nodes. Now we will setup our Linux Virtual Server (LBNODE1 & LBNODE2) by installing Piranha package. We already know that Piranha includes ipv-sadm, nanny and pulse demon. We will use Yum to install Piranha on the both nodes. Now we will configure Linux Virtual Server configuration file. Now we will copy this configuration file to lbnode2. Run this command on both nodes.

### 4.3 Install and Configure Apache (Web Server) Server

We will install Apache on both Nodes. Then start the httpd service on both web servers. We will start pulse service on both lbs nodes. Now we will install and configure our web servers and arptables_jf package for direct routing. Now we will configure the Ethernet interfaces for virtual IP on first web server node. Now we will do it on the second web server node. Now we will configure the Ethernet interfaces for virtual IP on second web server node.

Now we will configure our arptables on our first web server node. Now we will configure our arptables on our second web server node. We have managed to setup our LVS and webserver nodes now it's time to test if everything is working or not. Finally open a web browser from any machine and type http://www.orange.com and keep on refreshing page, we will get output of page contents from Webserver 1 and Web Server 2.

### 4.4 Install and Configure SAN (Openfiler)

Once the system boots up, start configuring Openfiler by pointing your browser at the host name or IP address of the Openfiler system. The interface is accessible from https port 446. e.g.. https://homer.the-simpsons.com:446 Management Interface: https://<ip of openfiler host>:446. Administrator Username: openfiler. Administrator Password: password

### 4.5 Testing using GUI

In the 3.3 (a-d) parts example, we find 1st request will handle websrv1 and 2nd request will be handle websrv2 and again 3rd request will handle websrv1 and process will be go on.

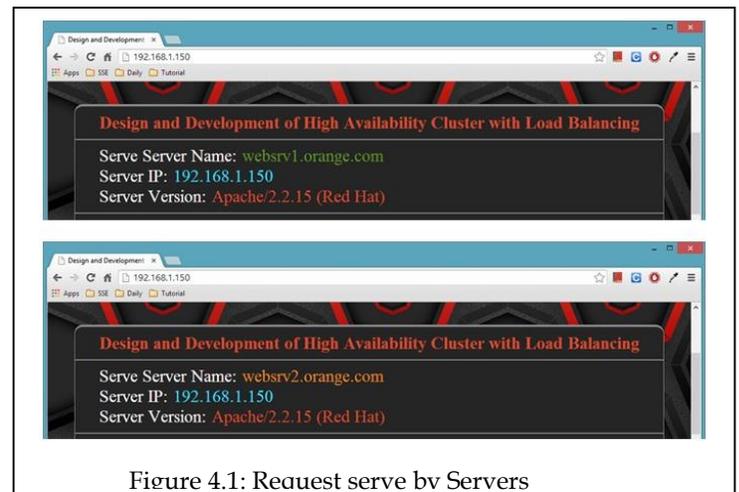

Figure 4.1: Request serve by Servers

Most important thing is all request will accept lbsrv1 until its goes down, when it will goes down then lbsrv2 will take over all the service and it will handle all the requests. Heartbeat network protocol is used this works to monitor. In the figure 4.1, with this high availability cluster model - first Request serve by websrv1.orange.com Server and second Request serve by websrv2.orange.com Server and so on.

### 4.6 Testing using Command Line

In the figure 4.2, we find 1st request will handle websrv1 and 2nd request will be handle websrv2 and again 3rd request will handle websrv1 and process will be go on.

Figure 4.2: PING Test (Request serve by Servers)

### 4.7 Perpormance Testing

In this section, we will test performance for this high availability cluster model. In the figure 4.3, we find LVS Cluster active on lbnode1.orange.com Node. Here 1st request will handle websrv1 and 2nd request will be handle websrv2 and again 3rd request will handle websrv1 and process will be go on.

Figure 4.3: LVS Cluster active on lbnode1.orange.com Node

### 4.8 Load Testing

In this section, we will test load for this high availability cluster model. In the figure 4.4, we find that for huge requests for process, LVS Cluster active on lbnode1.orange.com Node.

Figure 4.4: Load test - LVS Cluster active on lbnode1.orange.com Node

### 4.9 Failover Testing

In this section, we will test load for this high availability cluster model. In the figure 4.5, we find that for huge ping requests continuously for a certain time, LVS Cluster active on lbnode1.orange.com Node.

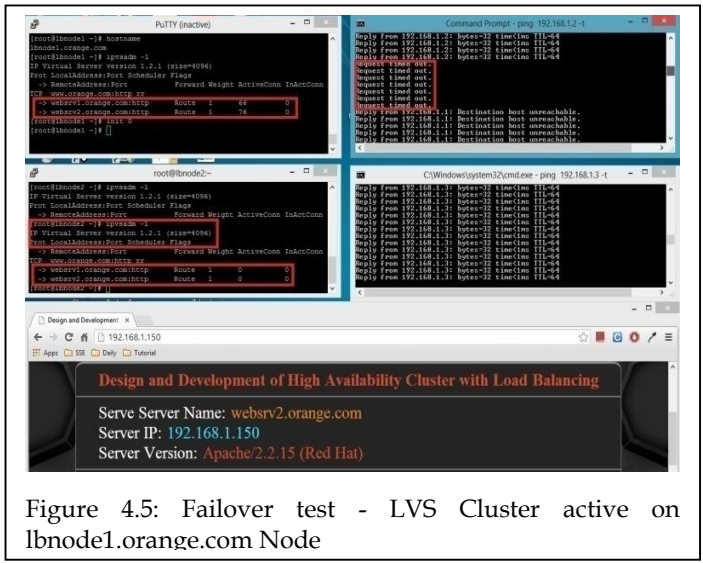

Figure 4.5: Failover test - LVS Cluster active on lbnode1.orange.com Node

## 6 Conclusion and Future Scope

This article focuses on how to build an LVS load-balancing cluster technology, combined with virtualization and shared storage technology to achieve the three-tier architecture of Web server clusters. This paper designs and implements a high availability clusters and incorporated with load balance infrastructure of web servers. The paper described system can provide full facilities to the website hosting provider and large business organizations.

In this paper explores experimental setup and conduct testing in the following: Testing using GUI, Testing using Command Line, Performance Testing, Load Testing, Failover Testing. This system can provide continuous service though any system components fail uncertainly with the help of Linux Virtual Server (LVS) load-balancing cluster technology and combined with virtualization as well as shared storage technology to achieve the three-tier architecture of Web server clusters. This technology not only improves availability, but also affects the security and performance of the application services being requested. Benefits of the system include node failover overcome; network failover overcome; storage limitation overcome and load distribution. This is a system which can provide full facilities to the website hosting provider and large business organizations High Availability Cluster with Load Balancing" is such a system, which can provide continuous service though any system components fail uncertainly. Without clustering, a server running a particular application as when that particular server crashes, the application will be unavailable until the crashed server is fixed. This system is very much efficient, because the full system is automated from the very beginning to the end.